\documentclass[final,5p,times,twocolumn]{elsarticle}

\usepackage{mypreamble}
\usepackage{ltablex}
\keepXColumns
\usepackage{color}
\usepackage{caption}
\usepackage{enumitem}
\setlist{nosep} 
\usepackage{array}
\usepackage{dblfloatfix}
\usepackage{pifont}

\usepackage{dsfont}
\usepackage{footnote}
\usepackage{tabularx}
\usepackage{threeparttable}
\usepackage{amsmath,amssymb,amsfonts}
\usepackage{amsmath}
\usepackage{multirow}
\usepackage[ruled,vlined]{algorithm2e}

\usepackage{hyperref}
\hypersetup{
    colorlinks=true,
    linkcolor=blue,
    filecolor=magenta,      
    urlcolor=cyan,
}
\usepackage{longtable}

\newacronym{ti}{$T_{i}$}{indoor air temperature, $^{\circ}$C}
\newacronym{tout}{$T_{out}$}{outdoor air temperature, $^{\circ}$C}
\newacronym{top}{$T_{o}$}{operative air temperature}
\newacronym{ts}{$T_{s}$}{supply air temperature, $^{\circ}$C}
\newacronym{v}{$\dot{V}$}{measured volumetric flow rate, L/s}
\newacronym{vsp}{$\dot{V}_{sp}$}{volumetric flow rate set-point, L/s}
\newacronym{tsph}{$T_{sp,h}$}{heating zone temperature set-point, $^{\circ}$C}
\newacronym{tspc}{$T_{sp,c}$}{cooling zone temperature set-point}
\newacronym{ghi}{$G_{hi}$}{global horizontal irradiance, W/m$^2$}
\newacronym{BMS}{BMS}{Building Management System}
\newacronym{pmv}{PMV}{Predicted Mean Vote}
\newacronym{HVAC}{HVAC}{Heating, Ventilation, and Air Conditioning}
\newacronym{VAV}{VAV}{Variable Air Volume}
\newacronym{AHU}{AHU}{Air Handling Unit}

\begin{document}


\begin{frontmatter}

\title{RRTS Dataset: A Benchmark Colonoscopy Dataset from Resource-Limited Settings for Computer-Aided Diagnosis Research}



\author[mymainaddress]{Ridoy Chandra Shil}
\author[mymainaddress]{Ragib Abid}
\author[mymainaddress]{Tasnia Binte Mamun}
\author[mymainaddress]{Samiul Based Shuvo}
\author[mysecondaryaddress4]{Dr. Masfique Ahmed Bhuiyan}
\author[mymainaddress]{Jahid Ferdous\corref{mycorrespondingauthor}}
\ead{ferdousj@bme.buet.ac.bd}



\cortext[mycorrespondingauthor]{Corresponding author \newline
Ridoy Chandra Shil, Ragib Abid and Jahid Ferdous are affiliated with: BioDesign Lab, Department of Biomedical Engineering, Bangladesh University of Engineering and Technology (BUET), Bangladesh
}


\address[mymainaddress]{Department of Biomedical Engineering, Bangladesh University of Engineering and Technology (BUET), Dhaka, Bangladesh}
\address[mysecondaryaddress4]{Dhaka Medical College, Dhaka, Bangladesh}


\begin{abstract}
\textbf{Background and Objective:}
Colorectal cancer prevention relies on early detection of polyps during colonoscopy. Existing public datasets, such as CVC-ClinicDB and Kvasir-SEG, provide valuable benchmarks but are limited by small sample sizes, curated image selection, or lack of real-world artifacts. There remains a need for datasets that capture the complexity of clinical practice, particularly in resource-constrained settings.

\noindent \textbf{Methods:}
We introduce a dataset, BUET Polyp Dataset (BPD), of colonoscopy images collected using Olympus 170 and Pentax i-Scan series endoscopes under routine clinical conditions. The dataset contains images with corresponding expert-annotated binary masks, reflecting diverse challenges such as motion blur, specular highlights, stool artifacts, blood, and low-light frames. Annotations were manually reviewed by clinical experts to ensure quality. To demonstrate baseline performance, we provide benchmark results for classification using VGG16, ResNet50, and InceptionV3, and for segmentation using UNet variants with VGG16, ResNet34, and InceptionV4 backbones.

\noindent \textbf{Results:}
The dataset comprises 1,288 images with polyps
from 164 patients with corresponding ground-truth masks and 1,657 polyp-free images from 31 patients. Benchmarking experiments achieved up to 90.8\% accuracy for binary classification (VGG16) and a maximum Dice score of 0.64 with InceptionV4-UNet for segmentation. Performance was lower compared to curated datasets, reflecting the real-world difficulty of images with artifacts and variable quality.

\noindent \textbf{Conclusions:}
This dataset provides a representative and challenging resource for developing robust computer-aided diagnosis systems in colonoscopy. By including diverse real-world imaging artifacts, it complements existing curated benchmarks and enables the development of models that generalize better to clinical practice. Baseline results are intended to guide future research and facilitate fair comparisons.
\noindent 
Please mail us to get the access of the dataset.
\end{abstract}


\end{frontmatter}

    \nolinenumbers



\section{Introduction} \label{mt1}

Colorectal cancer (CRC) is the third most common cancer globally, with approximately 1.9 million cases and over 900,000 deaths reported in 2020. Alarmingly, these numbers are expected to escalate, reaching 3.2 million new cases and 1.6 million deaths by 2040 \cite{morgan2023global}. Early diagnosis plays a crucial role in survival outcomes, with a 5-year survival rate of 91.5\% for Stage I CRC, dropping drastically to 16.2\% for late-stage diagnoses \cite{seer_colorectal}. Most CRC cases develop from precancerous polyps through the adenoma–carcinoma sequence. These statistics highlight the importance of early and accurate polyp detection. It is a key step in the prevention and timely treatment of CRC. Colonoscopy is widely regarded as the gold standard for CRC screening and polyp removal. It enables direct visualization of the entire colon and allows simultaneous biopsy and resection of suspicious lesions. However, despite its advantages, colonoscopy is not without limitations. Studies report that polyps can be missed in 9\% to 34\% of cases, with flat or sessile morphologies being particularly prone to oversight \cite{zhao2019magnitude}. The situation is even more critical in low-resource settings, where healthcare facilities often rely on older-generation equipment, limited technical expertise, and suboptimal imaging conditions. Hence, polyp detection accuracy is compromised.               

\begin{figure*}[!t]
    \centering
    \includegraphics[width=\textwidth]{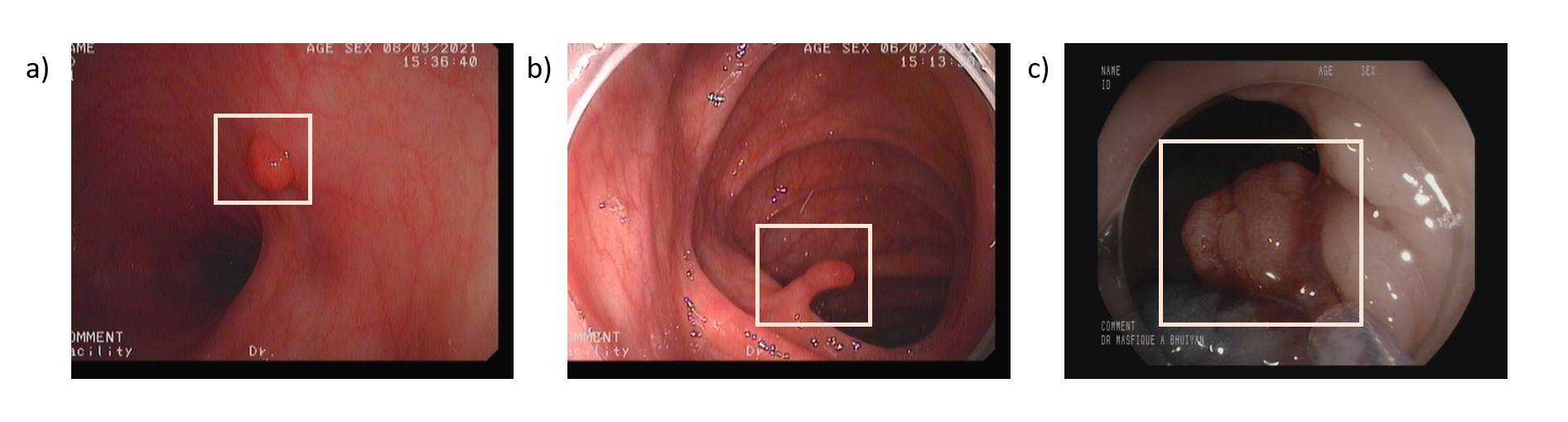}
    \caption{Examples of colorectal polyp morphologies observed during colonoscopy. (a) Sessile polyp lying flat against the colonic mucosa, making detection more challenging.(b, c) Pedunculated polyps with stalk-like structures protruding from the mucosal surface.}
    \label{fig:polyp}
\end{figure*}

\par
Recent advancements in artificial intelligence (AI), particularly deep learning–based computer-aided detection (CADe) systems, have shown promise in improving polyp detection rates and reducing inter-observer variability \cite{brown2022deep}. These systems rely heavily on the availability of high-quality, annotated datasets that capture the visual diversity and complexity of real-world colonoscopy images. While several public datasets such as Kvasir-SEG \cite{jha2019kvasir}, CVC-ClinicDB \cite{bernal2015wm}, ETIS-Larib \cite{silva2014toward}, and PolypGen \cite{ali2023multi} have contributed significantly to this domain, they suffer from several limitations. Many are curated from high-resource clinical environments, often using high-definition scopes. Images are pre-selected, typically excluding low-quality, blurry, or transitional frames. Dataset annotations may lack pixel-level precision, contain limited polyp morphology, or exclude non-polyp images. Several datasets are not publicly available, lack recommended data splits, or are highly imbalanced.

\par
To address these limitations, we present the BUET Polyp Dataset (BPD), a benchmark dataset comprising 2,945 colonoscopy images collected at a resource-constrained public hospital, Dhaka Medical College and Hospital in Bangladesh. Unlike existing datasets, BPD reflects the real-world clinical constraints encountered in under-resourced settings. It includes both polyp and non-polyp frames, and retains imperfections such as motion blur, glare, and variable illumination, which are typically excluded from curated datasets. Key contributions of BUET Polyp Dataset (BPD) dataset include:
\par 

\begin{enumerate}
    \item \textbf{Low-resource, real-world context:} Images are collected in a resource-limited public hospital, capturing challenges typical of such environments.

    \item \textbf{Segmentation-ready masks:} Each polyp image includes expert-verified, pixel-level binary masks, making the dataset suitable for training and evaluation of segmentation models.

    \item \textbf{Polyp and non-polyp cases:} Inclusion of negative samples enables robust training for binary classification tasks (polyp vs. non-polyp).
    
    \item \textbf{Generalization benchmark:} BPD serves as a stress test for AI models, highlighting performance gaps between curated and real-world environments. The dataset enables research in preprocessing, contrast enhancement, and artifact suppression algorithms designed for noisy, real-world inputs.
    
\end{enumerate}
 
The remainder of this paper is structured as follows: Section~2 provides the necessary background on colorectal cancer and polyps, highlighting their clinical significance. Section~3 reviews publicly available polyp datasets, emphasizing their characteristics and limitations. Section~4 describes the study design and data acquisition process, including subject selection, instrumentation, annotation strategy, and ethical considerations. Section~5 presents demographic and categorical analyses of the dataset. Section~6 details the overall data distribution and provides feature-level visualizations. Section~7 discusses challenges encountered during data preparation. Section~8 highlights the dataset’s impact and usability in real-world computer-aided diagnosis. Section~9 presents benchmarking and evaluation results for both segmentation and classification models. Section~10 provides a broader discussion of findings, followed by Section~11 on limitations, and finally, Section~12 concludes the paper with future directions.

    \section{Background} \label{mt2}
\subsection{Colorectal Cancer and Polyps}

\begin{figure}[!h]
    \centering
    \includegraphics[width=\columnwidth]{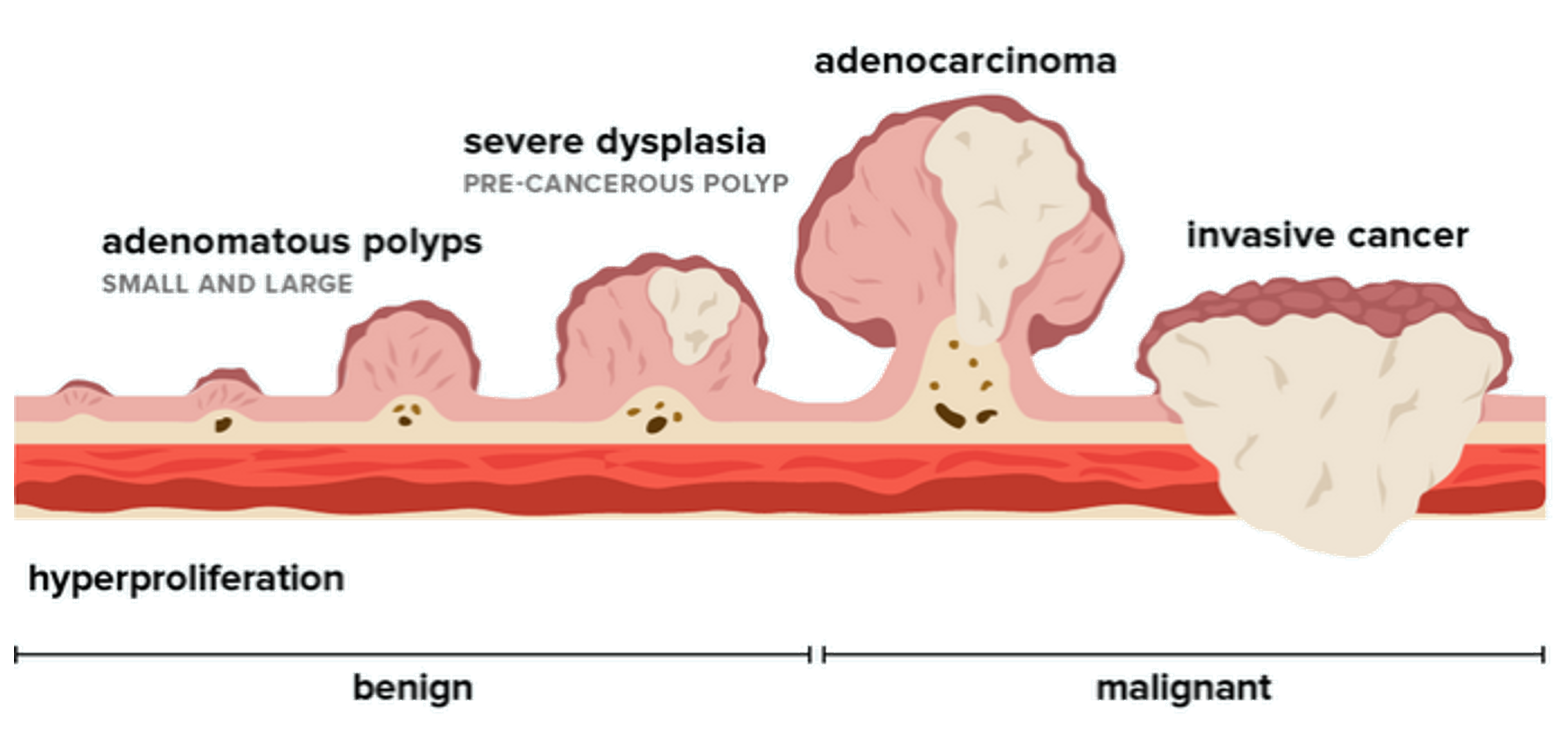}
    \caption{Progression of colorectal cancer from benign adenomatous polyps to malignant invasive cancer\cite{healthline_polyp_size}.}
    \label{fig:size}
\end{figure}

CRC typically originates from the benign overgrowth of mucosal epithelial cells. These lesions, termed polyps, may persist and enlarge slowly over a period of 10–20 years before undergoing malignant transformation \cite{rawla2019epidemiology}. The most frequent type is the adenomatous polyp, which arises from glandular cells responsible for secreting mucus within the large intestine \cite{stryker1987natural}. Although only around 10\% of adenomas ultimately develop into invasive carcinoma, the probability of malignant change increases with polyp size as shown in Fig.~\ref{fig:size}. When such polyps progress to invasive disease, the resulting malignancy is termed adenocarcinoma, which accounts for approximately 96\% of all CRC cases \cite{stewart2006population}.

\par
Table\ref{tab:polyp-types} gives an overview of the five most common types of polyps, their cancer risks, and treatment. 

\begin{table*}[t]
\centering
\caption{Comparison of Common Polyp Types Based on Prevalence, Cancer Risk, and Treatment}
\label{tab:polyp-types}
\renewcommand{\arraystretch}{1.5}
\resizebox{\linewidth}{!}{
\begin{tabular}{p{2.5cm}|p{5cm}|p{3cm}|p{5cm}}
\hline
\hline
\multicolumn{1}{c|}{\textbf{Type of polyp}} & 
\multicolumn{1}{c|}{\textbf{Prevalence}} & 
\multicolumn{1}{c|}{\textbf{Cancer Risk}} & 
\multicolumn{1}{c}{\textbf{Treatment}} \\
\hline
\hline
Inflammatory & Frequently observed in patients with chronic inflammatory conditions like Crohn’s disease or ulcerative colitis & Generally non-cancerous; low malignant potential & Typically removed during colonoscopy \\
\hline
Hyperplastic & Commonly appear near the rectum or end portion of the colon and are usually small & Regarded as low-risk for cancer & Polypectomy is usually performed during colonoscopy \\
\hline
Adenomatous (Tubular Adenoma) & Most prevalent variant; responsible for approximately 70\% of all colorectal polyps & Majority remain benign, though larger ones carry an increased risk of progressing to cancer & Excised during colonoscopy; routine monitoring may be required to detect recurrence \\
\hline
Villous or Tubulovillous Adenoma & Account for around 15\% of all polyps found during colonoscopy & While many remain noncancerous, larger lesions have greater potential for malignancy & Can be difficult to extract due to flat shape; smaller ones are usually removed endoscopically, but larger ones might need surgical intervention \\
\hline
Serrated Adenoma & Represent about 10–15\% of polyp cases & Associated with 20–30\% of colorectal cancers & Detection is challenging during colonoscopy due to subtle appearance \\
\hline
\hline
\end{tabular}}
\end{table*}

Colorectal polyps generally exhibit two distinct morphologies: sessile (flat) and pedunculated (stalked) as shown in Fig.\ref{fig:polyp}. Sessile polyps, which are now recognized as more prevalent than once believed, are particularly challenging to detect during colorectal cancer screening due to their flat structure that closely adheres to the mucosal surface of the colon. In contrast, pedunculated polyps resemble a mushroom-like structure, protruding from the mucosal lining and connected by a narrow, elongated stalk \cite{uofm_colon_polyps} .
\subsection{Colonoscopy}
Colonoscopy is the gold standard for colorectal cancer screening, recommended every 10 years for average-risk individuals over 50. Early detection and removal of polyps during colonoscopy significantly reduces morbidity and mortality. It achieves >95\% sensitivity for CRC and 88–98\% for advanced adenomas, and observational studies suggest it reduces CRC incidence by up to 72\% and mortality by about 31\%. Its major advantage is the ability to both detect and remove precancerous lesions in a single procedure \cite{simon2016colorectal}. Colonoscopy is conducted using a flexible, hand-held instrument known as a colonoscope, which is equipped with a high-definition camera at its distal end. The device also contains accessory channels that facilitate the passage of instruments and fluids used to clean both the lens and the colonic lining. The real-time video feed transmitted by the camera enables clinicians to identify mucosal abnormalities, including overgrowths of the colonic wall. This visual guidance allows for the evaluation, biopsy, and removal of lesions using various biopsy tools introduced through the accessory channels. Owing to its wide-ranging diagnostic and therapeutic capabilities, colonoscopy has become a cornerstone in the early detection and prevention of colorectal cancer over recent decades.

    \section{Available Polyp Datasets} \label{mt3}
A wide range of benchmark datasets have been developed to support the training and evaluation of computer-aided detection (CADe) and computer-aided diagnosis (CADx) systems in colonoscopy. Among them, a few have gained particular prominence due to their accessibility, size, and influence on benchmarking studies. 

\subsection{CVC-ClinicDB} 
CVC-ClinicDB~\cite{bernal2015wm} is one of the earliest and most widely used benchmark datasets for polyp segmentation. It consists of 612 still frames extracted from 31 colonoscopy videos collected at Hospital Clinic, Barcelona, Spain. Each frame is provided with binary ground-truth masks annotated by expert endoscopists, marking the polyp regions at the pixel level. Although relatively small in size, CVC-ClinicDB has become a standard reference point for segmentation methods due to its accessibility and consistent annotations. Many early deep learning works used this dataset for both training and evaluation, often in combination with CVC-ColonDB to mitigate overfitting risks. Its main limitation lies in the small cohort and the fact that only polyp-containing frames are included, without negative samples for model robustness.
\begin{table*}[!h]
\centering
\caption{Benchmark datasets for polyp detection and segmentation.}

\label{tab:polyp_datasets1}
\begin{normalsize}

\setlength{\tabcolsep}{4pt}
\renewcommand{\arraystretch}{1}
\begin{tabular}{p{0.5cm}|p{2cm}|p{1cm}|p{.95cm}|p{1.5cm}|p{1.35cm}|p{1.75cm}|p{4cm}}
\hline
\hline
\multicolumn{1}{c|}{\textbf{Sl.}} & 
\multicolumn{1}{c|}{\textbf{Dataset}} & 
\multicolumn{1}{c|}{\textbf{No. of Patients}} & 
\multicolumn{1}{c|}{\textbf{Type}} & 
\multicolumn{1}{c|}{\textbf{Samples}} & 
\multicolumn{1}{c|}{\textbf{Ground Truth}} & 
\multicolumn{1}{c|}{\textbf{Acquisition Site}} & 
\multicolumn{1}{c}{\textbf{Limitations}} \\
\hline

\hline
1 & ASU-Mayo Clinic Colonoscopy video database\cite{tajbakhsh_automated_2016} & – & Images & 19,400 (5,200 polyp, 14,200 normal) & Binary mask & USA & Only 10 unique polyps; class imbalance; test set unavailable; motion-blurred frames unannotated; access on request. \\
\hline
2 & GI-Lesion\cite{mesejo_computer-aided_2016} & – & Videos & 76 & Coarse ROI & France & No pixel-level annotations; no negative samples; imbalanced distribution; coarse ROI unsuitable for segmentation. \\
\hline
3 & CVC-EndoScene-Still\cite{vazquez_benchmark_2017} & 36 & Images & 912 & Multi-class mask (4) & Spain & Not publicly available; small dataset; low polyp diversity. \\
\hline
4 & NBI-UCdb\cite{figueiredo_unsupervised_2019} & 10 & Images & 86 (11 videos) & Binary mask & Portugal & Not publicly available; very small dataset; imbalanced classes; overfitting risk. \\
\hline
5 & KUMC\cite{patel_comparative_2020} & – & Images & 37,899 & Bounding box & USA & No segmentation masks; many frames from same polyp; negative samples underrepresented. \\
\hline
6 & SUN\cite{misawa_development_2021,itoh_sun_2020} & 99 & Images & 158,690 & Bounding box & Japan & No segmentation masks; restricted access; many redundant frames. \\
\hline
7 & PICCOLO\cite{sanchez-peralta_piccolo_2020} & 48 & Images & 3,433 & Binary mask & Spain & Approval required; limited dataset size; all polyps centered (bias risk). \\
\hline
8 & ClinExpPIC-COLO\cite{sanchez-peralta_clinical_2023} & – & Images & 65 & Binary mask & Spain & Access on request; extremely small dataset. \\
\hline
9 & LD-Polyp-Video\cite{ma_ldpolypvideo_2021} & – & Images & 901,666 (40k annotated) & Bounding box & China & No pixel-level masks; labeling errors and loose bounding boxes reported\cite{wei2022boxpolyp}. \\
\hline
10 & SUN-SEG\cite{ji_video_2022} & – & Images + Videos & 1,106 videos, 158k frames & Labels, masks, bounding boxes, polygons & Japan & Access on request; no standard split; class imbalance. \\
\hline
11 & NeoUNet\cite{bebis_neounet_2021} & – & Images & 7,466 & Multi-class mask (4 classes) & Vietnam & Severe class imbalance. \\
\hline
12 & EDD2020\cite{ali_endoscopy_2020} & 137 & Images & 386 & Binary mask + bounding box & Europe & Small dataset; imbalanced classes; approval required. \\
\hline
13 & CVC-Clinic-VideoDB\cite{angermann_towards_2017} & – & Videos + Frames & 18 videos, 10,924 frames & Binary mask & Spain & Limited polyp diversity; ellipse-based annotations are imprecise. \\
\hline
14 & HyperKvasir \cite{borgli2020hyperkvasir} & – & Images & 1000 & Binary mask & Spain & Domain shift; expert selected images, missing challenging cases \\
\hline
\hline
\end{tabular}
\end{normalsize}
\end{table*}

\begin{table*}[!h]
\centering
\caption{Benchmark datasets for polyp classification.}

\label{tab:polyp_classification_datasets}
\begin{normalsize}
\setlength{\tabcolsep}{4pt}
\renewcommand{\arraystretch}{1}
\begin{tabular}{p{0.5cm}|p{2cm}|p{1cm}|p{.95cm}|p{1.5cm}|p{1.35cm}|p{1.75cm}|p{4cm}}
\hline
\hline
\multicolumn{1}{c|}{\textbf{Sl.}} & 
\multicolumn{1}{c|}{\textbf{Dataset}} & 
\multicolumn{1}{c|}{\textbf{No. of Patients}} & 
\multicolumn{1}{c|}{\textbf{Type}} & 
\multicolumn{1}{c|}{\textbf{Samples}} & 
\multicolumn{1}{c|}{\textbf{Ground Truth}} & 
\multicolumn{1}{c|}{\textbf{Acquisition Site}} & 
\multicolumn{1}{c}{\textbf{Limitations}} \\
\hline

\hline
1 & WL-UCdb\cite{figueiredo_polyp_2019,figueiredo_fast_2020} & 42 & Images & 3,040 (1,680 polyp, 1,360 normal) & File-level binary classification label & Portugal & Not publicly available; duplicate-like images; possible data leakage. \\
\hline
2 & CP-CHILD\cite{wang_improved_2020} & 1600 & Images & 9,500 & File level binary classification label & China & Class imbalance; pediatric data only; blurry/obscured frames excluded. \\
\hline
3 & ERCPMP\cite{forootan_ercpmp_2024} & 191 & Images + Videos & 796 images, 21 videos & Classification labels & Iran & Class imbalance. \\
\hline
4 & REAL-Colon\cite{biffi2024real} & 60 & Videos + Frames & 60 videos, 2.7 million images & Bounding box, histological diagnosis & Japan, Austria, Italy, USA & Limited cohort size; lower quality data removed \\
\hline
5 & HyperKvasir \cite{borgli2020hyperkvasir} & - & Images and videos & 110k images (10k labeled), 374 videos (all labeled) & Class labels, segmentation mask & Norway & class imbalance; majority data unlabeled; potential duplicates; domain shift issues \\
\hline
\hline
\end{tabular}
\end{normalsize}
\end{table*}

\subsection{ETIS-Larib} 
ETIS-Larib~\cite{silva2014toward} is another influential dataset, designed to evaluate the generalization capability of polyp detection and segmentation models. It comprises 196 frames extracted from 34 sequences collected at the Lariboisière Hospital, Paris, France. Polyps in this dataset are generally small, subtle, and frequently subject to challenging imaging conditions such as specular highlights, motion blur, and poor contrast. The ground-truth masks were annotated by clinical experts. Due to its difficulty and relatively small size, ETIS-Larib is rarely used as a training dataset; instead, it serves as a “hard test set” to benchmark the robustness of models trained on datasets like CVC-ClinicDB or Kvasir-SEG.

\subsection{Kvasir}
Kvasir~\cite{pogorelov2017kvasir} is a popular multi‐class image dataset comprising of images from the gastrointestinal (GI) tract collected at Vestre Viken Health Trust, Norway, using standard endoscopic imaging equipment. Annotation was done by experienced endoscopists and experts from the Cancer Registry of Norway. The dataset contains 4,000 images divided evenly into eight classes (about 500 per class), with each class representing either anatomical landmarks, pathological findings, or procedures related to polyp removal. Image resolution is variable, ranging from about 720×576 up to 1920×1072 pixels organized into folders by class. Because of its multi‑class nature, relatively balanced class sizes, good image quality, and diversity in anatomical regions and pathologies, Kvasir is widely used for both classification and detection tasks. Overall, Kvasir occupies an important place as a middle‑scale, well‐annotated, multi‑class GI image dataset: large enough to enable many machine learning experiments, yet manageable and well understood.

\subsection{Kvasir-SEG} 
Kvasir-SEG~\cite{jha2019kvasir} is one of the most popular open-access datasets for polyp segmentation, developed by Simula Research Laboratory and the Cancer Registry of Norway. It contains 1,000 polyp images, each paired with pixel-level ground-truth masks and corresponding bounding boxes. Images were acquired under diverse imaging conditions and exhibit considerable variation in polyp size, shape, and morphology. Unlike earlier datasets, Kvasir-SEG includes both easy and difficult cases, such as flat polyps and images with partial occlusions, which makes it highly suitable for benchmarking modern deep learning models. Its public availability, standardized annotations, and balance between dataset size and diversity have made it one of the most widely adopted datasets in recent years.

\subsection{CVC-ColonDB} 
CVC-ColonDB~\cite{bernal_towards_2012} is an earlier dataset produced at the same institution as CVC-ClinicDB. It comprises 300 frames extracted from 15 colonoscopy videos, with pixel-level binary masks provided for polyp segmentation. Although modest in size, CVC-ColonDB remains relevant in the literature, especially when paired with CVC-ClinicDB to form a combined training and testing resource. However, it lacks negative samples and contains fewer variations in polyp morphology and imaging conditions compared to Kvasir-SEG. Nonetheless, it played an important role in establishing early benchmarks for polyp segmentation and continues to be cited in comparative studies.

\subsection{PolypGen} 
PolypGen~\cite{ali2023multi} represents the latest effort toward creating a large-scale, multi-center dataset for robust polyp detection and segmentation. It contains 8,037 images from 300 patients, annotated with both pixel-level binary masks and bounding boxes. A major strength of PolypGen lies in its multi-institutional and multi-country acquisition strategy, covering diverse patient populations and endoscopy systems. This diversity makes it far more representative of real-world clinical conditions compared to earlier datasets, which were often limited to single centers. PolypGen also includes both polyp and non-polyp frames, supporting both classification and segmentation tasks. With its scale, diversity, and clinical realism, it currently stands as the most comprehensive publicly available dataset for polyp-related computer vision research.

Beyond these, a number of additional datasets exist, covering different acquisition settings, annotation schemes, and clinical focuses. A comprehensive summary of these resources is provided in Table~\ref{tab:polyp_datasets1}. Additionally, Table \ref{tab:polyp_classification_datasets} summarizes some important datasets for polyp classification.

    \section{Study Design and Data Acquisition} \label{mt4}

The absence of strong and robust computer-assisted tools for polyp detection and segmentation in colonoscopy images in the lower resource setting motivated the creation of this dataset. Colonoscopy image data were retrospectively collected from routine clinical procedures at Dhaka Medical College and Hospital (DMCH), Dhaka, Bangladesh. Both polyp and non-polyp images were included to ensure that the dataset reflects the real clinical distribution of cases and can be used for both classification and segmentation tasks. The data was annotated by trained annotators and reviewed by expert gastroenterology surgeon. To ensure stable model evaluation, the data were split at the patient level such that images from the same patient would not be present in both test and training sets. Finally, all data were anonymized and renamed with UUIDs to completely remove patient identifiers and preserve privacy. This carefully crafted dataset is a valuable asset towards advancing computer-aided polyp detection research for resource-constrained clinical settings.

\subsection{Study Procedure}

The research process involved retrospective collection of colonoscopy images from multiple patients at Dhaka Medical College and Hospital. All the images were anonymized in their entirety without access to clinical or demographic information. The pipeline began with the removal of low-quality, incomplete, or inappropriate images by an initial review. Next, polyp regions were localized and annotated by applying bounding boxes followed by precise polygonal segmentation masks. Multiple passes of expert validation were conducted to verify the accuracy of the annotations. A strict patient-level split was performed to prevent source data contamination. After that universally unique identifiers (UUIDs) were assigned to all images to prevent any potential traceability of patients.
  


\subsection{Hardware Setup}
Colonoscopy data were acquired using Olympus 170 series colonoscopes and Pentax i-Scan series endoscopy systems. The Olympus 170 provides standard white-light imaging, while the Pentax i-Scan platform employs digital image enhancement with surface, tone, and contrast modes. These devices were used during routine clinical procedures to generate the dataset analyzed in this study.

\subsection{Subject Inclusion Criteria}

Subjects were randomly selected from a large pool based on the availability of colonoscopy images suitable for polyp detection and segmentation. Only patients whose images were clear enough for reliable annotation were included, without consideration of age, gender, or medical history, as all personal information had been removed.

\subsection{Ethical Considerations}
This study was conducted using fully anonymized, retrospective colonoscopy images collected as part of routine clinical care at Dhaka Medical College and Hospital. No identifiable patient information was used. In accordance with institutional and national guidelines, formal ethics committee approval and informed consent were not required for this type of secondary data analysis.

\subsection{Data Annotation Strategy}


Creating reliable annotations was a core part of this task, and significant effort was dedicated to precision and consistency within the annotation process. The process followed the following multi step approaches:
\begin{itemize}
  \item Initial Localization: Annotators used the labelImg\cite{labelImg} package to sketch bounding boxes over visible polyps, establishing a primary region of interest.

  \item Fine-Grained Segmentation: With the bounding boxes as reference, labelme\cite{Wada_Labelme_2021} tool was used to create highly detailed polygonal masks. As a result polyp boundaries were captured with improved precision.

  \item Expert Verification: All the annotations were verified by expert gastroenterology surgeon through multiple validation steps. Images with unsatisfactory and/or ambiguous annotations were re-verified, revised, or deleted.
\end{itemize}
This careful, multi-phase approach was designed to minimize errors, normalize annotations, and produce a dataset reflective of polyp clinical appearance so it can be valuable both for training and for evaluation.
    
    \section{Demographic Analysis of Polyp Dataset}

The polyp dataset used in this research was constructed retrospectively from a large collection of colonoscopy images taken from multiple patients. As this was a secondary use of existing data, all patient information was fully anonymized at the time of data preparation. We had no access to demographic information such as age, sex, or clinical history, and patients were only identified by unique identifiers. In order to prevent data leakage and enforce a rigorous evaluation framework, we performed patient-level splitting of the both segmentation and classification dataset prior to training the models in such a way that images of any given patient were present solely in either the training or testing subset. As an additional measure of privacy, all patient identifiers were replaced with randomly generated UUIDs after splitting, further removing any links to the original cases. This random selection and anonymization process yielded a final dataset of 1,288 polyp images from 164 patients for segmentation and 2,945 images from 195 patients for classification, constituting a large and privacy-preserved dataset for AI model development.








    \section{Categorical Data Analysis}
Colonoscopy image datasets are inherently heterogeneous, containing frames acquired under a variety of clinical conditions. As illustrated in Table~\ref{tab:polyp_examples_part1}, the data frequently exhibit artifacts that complicate polyp detection and segmentation. Examples include stool residues adhering to the intestinal wall, blood obscuring mucosal surfaces, or inflammation that alters tissue appearance. Additional challenges arise from image acquisition itself, such as low brightness, chromatic aberration, motion blur, or the presence of surgical tools in the field of view.

\begin{table*}[!h]
\centering
\caption{Challenging Polyp image examples of BPD dataset.}
\label{tab:polyp_examples_part1}
\begin{normalsize}
\begin{tabular}{c|>{\raggedright\arraybackslash}m{7cm}|m{3cm}|m{3cm}}

\hline
\hline
\multicolumn{1}{c|}{\textbf{Sl. no}} & 
\multicolumn{1}{c|}{\textbf{Image description}} & 
\multicolumn{1}{c|}{\textbf{Image}} & 
\multicolumn{1}{c}{\textbf{Mask}} \\
\hline
\hline
1 & Image containing polyp without any noise or challenging aspects & \includegraphics[width=2.5cm]{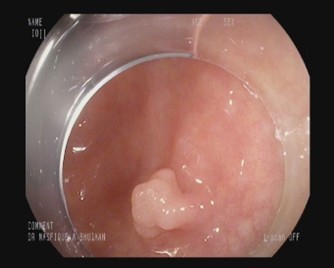} & \includegraphics[width=2.5cm]{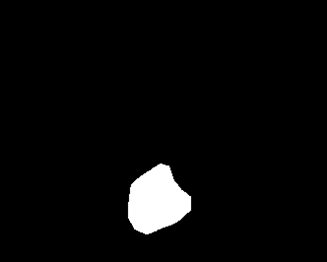} \\
\hline
2 & Image contains polyp with some stool dotting the intestinal wall & \includegraphics[width=2.5cm]{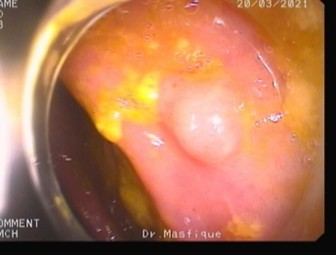} & \includegraphics[width=2.5cm]{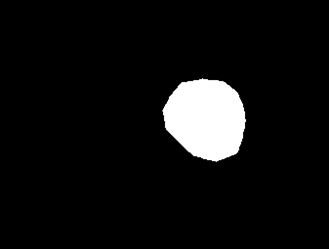} \\
\hline
3 & Image contains polyp with blood being present on the intestine wall & \includegraphics[width=2.5cm]{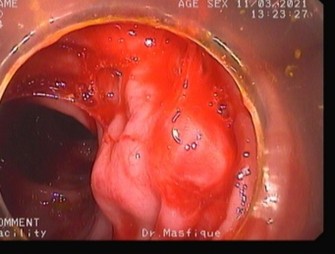} & \includegraphics[width=2.5cm]{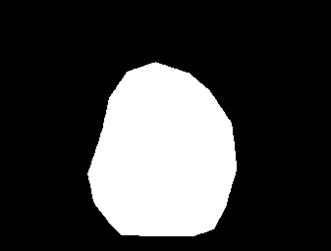} \\
\hline
4 & Image contains polyp in an inflamed intestine & \includegraphics[width=2.5cm]{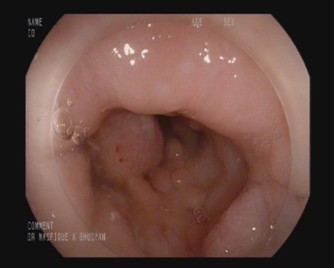} & \includegraphics[width=2.5cm]{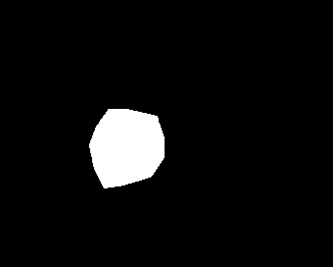} \\
\hline
5 & Blurred or unfocused image containing polyp & \includegraphics[width=2.5cm]{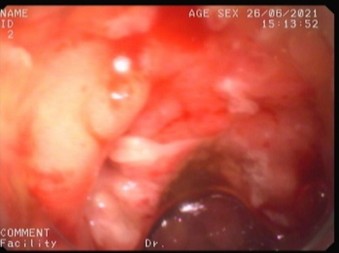} & \includegraphics[width=2.5cm]{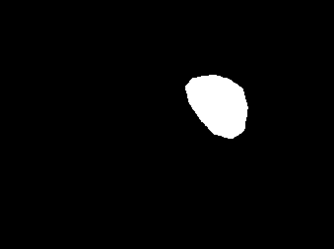} \\
\hline

6 & Chromatic aberration distortion present in polyp image & \includegraphics[width=2.5cm]{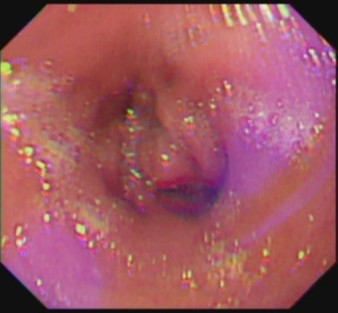} & \includegraphics[width=2.5cm]{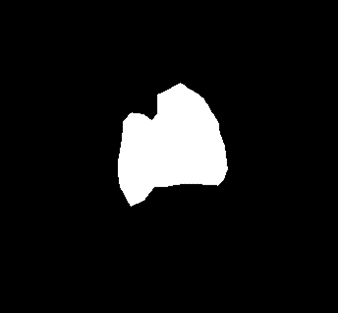} \\
\hline
7 & Low brightness image containing polyp & \includegraphics[width=2.5cm]{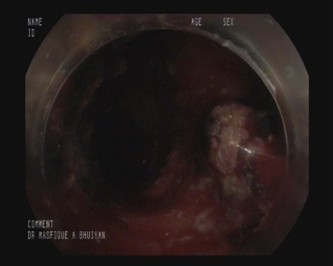} & \includegraphics[width=2.5cm]{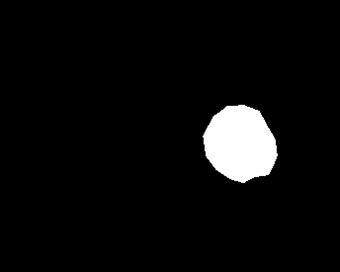} \\
\hline
8 & Image with improper bowel preparation & \includegraphics[width=2.5cm]{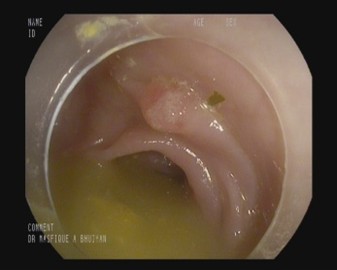} & \includegraphics[width=2.5cm]{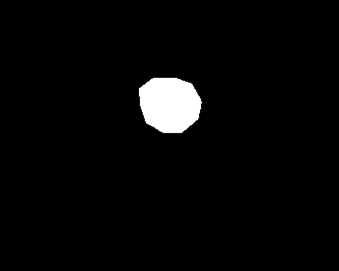} \\
\hline
9 & Surgical tool present alongside polyp & \includegraphics[width=2.5cm]{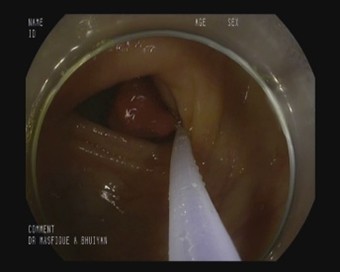} & \includegraphics[width=2.5cm]{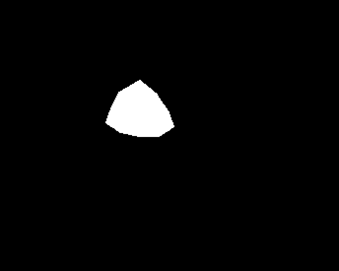} \\
\hline
10 & Motion blur present in image & \includegraphics[width=2.5cm]{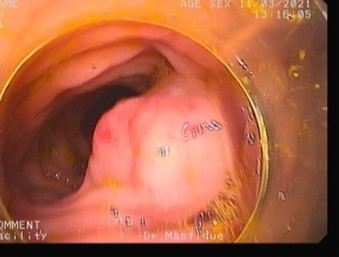} & \includegraphics[width=2.5cm]{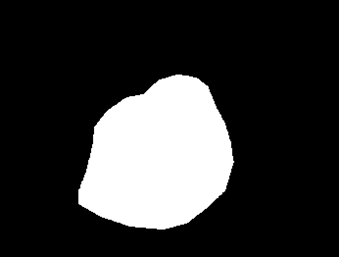} \\
\hline
\hline
\end{tabular}
\end{normalsize}
\end{table*}

    \section{Data Distribution}


The segmentation dataset contains 1,288 colonoscopy images from 164 unique patients. All the images have been carefully annotated by trained annotators to specify the areas in which the polyps exist at the pixel level for precise semantic segmentation. The dataset was divided at the patient level to prevent data leakage and enhance generalization of trained models. A total of 1,032 images of 135 patients were used for model training and the other 256 images of 29 patients for independent testing only. The division guarantees that images of the same patient are not present both in the training set and the test set to ensure an overall evaluation of model performance.

The classification dataset is derived from the same cohort of patients and includes a total of 2,945 colonoscopy images of 195 patients. The dataset includes 1,288 images with polyps from 164 patients and 1,657 polyp-free images from 31 patients, thereby including a large variety of pathological and normal cases. To train the classification models, 2,355 images (1,032 polyp and 1,323 non-polyp) from 161 patients were used for training, and a test set of 590 images (256 polyp and 334 non-polyp) from 34 patients was reserved for the final testing. This distribution of the dataset makes it representative of natural class balance between both splits and allows unbiased benchmarking of polyp detection algorithms.

\begin{figure}[!t]
    \centering
    \includegraphics[width=1.0\linewidth]{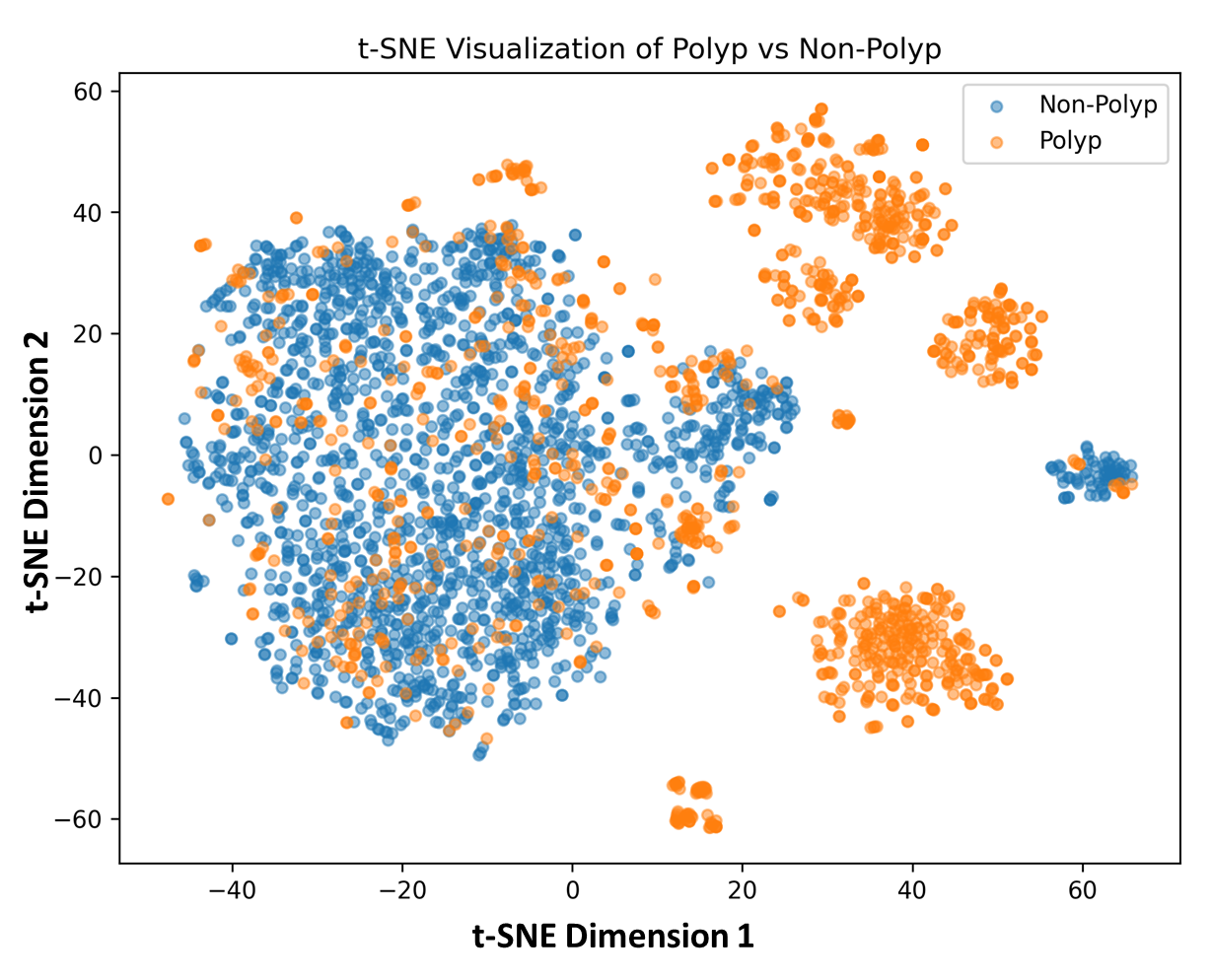}
\caption{t-SNE visualization of ResNet18 feature embeddings for polyp and non-polyp images. Polyp samples form several compact clusters, while non-polyp samples are more diffusely distributed, with notable overlap between the two classes.}

    \label{Qual1}
\end{figure}

Fig.~\ref{Qual1} shows the t-SNE projection of feature embeddings for polyp and non-polyp images. Several compact clusters dominated by polyp samples indicate discriminative visual cues, whereas non-polyp samples form a broader, more diffuse distribution. A central region of overlap highlights visual similarities between certain polyps and normal mucosa, suggesting the need for task-specific fine-tuning to enhance separability. Overall, the embedding demonstrates that pretrained CNN features capture meaningful structure but still exhibit inter-class overlap.

    \section{Challenges in Data Preparation}

Preparation of this dataset was subject to various challenges, primarily because of its origin in a low-resource clinical setting. Images were viewed and annotated by trained personnel, but it was difficult to achieve absolute and perfectly uniform annotations in the presence of variability in image quality, light intensity, bowel preparation quality, and polyp appearance. Extensive labeling time was constrained by resources, and ambiguous cases occasionally required multiple rounds of review. These conditions introduced inconsistencies in ensuring uniform quality of annotations across the dataset and may have led to under-annotation or over-annotation of polyp regions on certain instances. Despite these drawbacks, strict curation and quality control by the professionals were imposed for ensuring maximum accuracy in annotations and consistency in the dataset.

    \section{Impact and Usability}
The introduction of this colonoscopy dataset carries considerable impact for healthcare and research, particularly in resource-constrained environments. By incorporating both polyp images with expert-verified segmentation masks and non-polyp images, the dataset supports not only pixel-level segmentation tasks but also classification studies. This dual usability expands its relevance for developing comprehensive computer-aided diagnosis (CADx) pipelines, enabling both lesion localization and polyp presence detection.  

Unlike curated datasets from high-resource clinical settings, this collection reflects real-world challenges such as glare, motion blur, and uneven illumination, making it highly valuable for building AI systems robust to practical imaging artifacts. In underdeveloped and developing countries, where access to high-definition endoscopy equipment and trained gastroenterologists remains limited, such a dataset provides an essential resource for improving colorectal cancer screening and strengthens global AI research by addressing the underrepresentation of low-resource clinical conditions in existing benchmarks. It can aid in reducing polyp miss rates through reliable classification models and enhance lesion delineation via segmentation networks, thereby facilitating earlier diagnosis and intervention.  

In conclusion, this dataset not only advances methodological research in medical image analysis but also contributes to reducing disparities in cancer care between high- and low-resource regions.

\section{Benchmarking and Evaluation}
\begin{figure}[!t]
    \centering
    \includegraphics[width=1.0\linewidth]{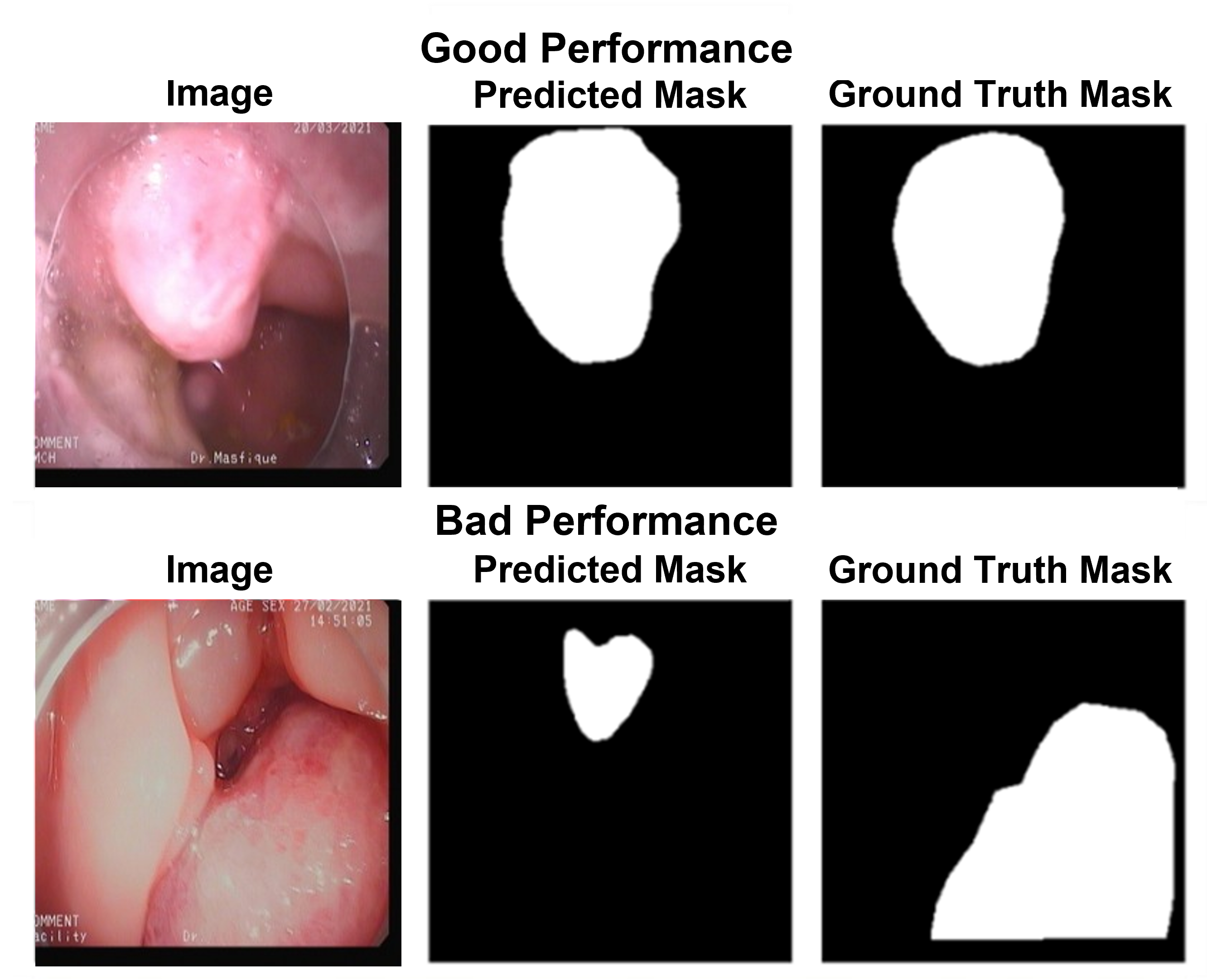}
    \caption{Qualitative examples of segmentation performance of InceptionV4-UNet model. The top row shows a case of good performance where the predicted mask closely matches the ground truth. The bottom row shows a case of poor performance, where the prediction fails to capture the polyp boundaries accurately.}
    \label{Qual}
\end{figure}
\subsection{Model Architectures}
Two complementary tasks were considered: (i) polyp segmentation and (ii) binary classification of polyp versus non-polyp frames.  

\subsubsection{Segmentation models:}  
We benchmarked four UNet-based architectures: baseline UNet, VGG16-UNet, ResNet34-UNet, and InceptionV4-UNet. The baseline UNet follows a canonical encoder–decoder design, where convolutional and pooling layers extract features that are progressively upsampled through the decoder, with skip connections preserving spatial resolution. In the extended variants, the encoder is replaced with pretrained ImageNet backbones (VGG16, ResNet34, InceptionV4), allowing the models to leverage transfer learning for more robust feature extraction. The decoder layers remain trainable and reconstruct segmentation masks from the enriched encoder representations. A sigmoid-activated output layer generates binary masks.  

\subsubsection{Classification models:}  
For binary classification of polyp versus non-polyp images, three widely used CNNs VGG16, ResNet34, and InceptionV4 were fine-tuned using transfer learning. In all three models, the fully connected classifier head was modified to include dropout layers, followed by a single sigmoid-activated neuron for binary prediction.  

\subsection{Training Methodology}
For segmentation, dataset was divided into training and test sets using a patient-wise split (1,032 training images, 256 test images) to prevent data leakage. For classification, balanced subsets of polyp and non-polyp images were used. Data augmentation (rotations, flips, brightness/contrast shifts) was applied to increase robustness. All models were trained using Adam ($\alpha = 1\times10^{-4}$) with binary cross-entropy loss, batch size 8 (segmentation) or 32 (classification), and early stopping to prevent overfitting.  
\begingroup
\renewcommand\thefootnote{}\footnote{The code is publicly available at:~\href{https://github.com/RidoyChandraShil/BUET_Polyp_Dataset}{BUET Polyp Dataset (BPD) code Access Link}}%
\addtocounter{footnote}{-1}%
\endgroup
\subsection{Evaluation Procedure}
To assess the performance of the models, both segmentation and classification metrics were computed on the test set.  

For segmentation, two widely adopted overlap-based metrics were used: the Dice coefficient and the Intersection over Union (IoU). These are defined as:  

\begin{equation} 
\text{Dice} = \frac{2|P \cap G|}{|P| + |G|} 
\end{equation}

\begin{equation} 
\text{IoU} = \frac{|P \cap G|}{|P \cup G|} 
\end{equation}

where $P$ and $G$ represent the sets of predicted and ground-truth pixels, respectively. Dice reflects the overall overlap quality, while IoU provides a stricter penalty for mismatched regions.  

For binary classification of polyp versus non-polyp images, the following standard metrics were used: accuracy, sensitivity, specificity, and the F1 score. These were derived from the confusion matrix terms: True Positives (TP), True Negatives (TN), False Positives (FP), and False Negatives (FN).  

\begin{equation} 
\text{Accuracy (Acc.)} = \frac{TP + TN}{TP + TN + FP + FN} 
\end{equation}

\begin{equation} 
\text{Sensitivity (Sen.)} = \frac{TP}{TP + FN} 
\end{equation}

\begin{equation} 
\text{Specificity (Spec.)} = \frac{TN}{TN + FP} 
\end{equation}

\begin{equation} 
\text{F1 Score} = \frac{2TP}{2TP + FP + FN} 
\end{equation}

Accuracy measures overall correctness, sensitivity quantifies the proportion of actual polyps correctly detected, and specificity reflects the proportion of non-polyp cases correctly identified. The F1 score balances false positives and false negatives, making it suitable in scenarios where class imbalance may occur. Together, these metrics provide a comprehensive evaluation of both spatial segmentation accuracy and image-level classification performance.

\subsection{Results}
\subsubsection{Segmentation}  
Table~\ref{tab:segmentation_results} summarizes segmentation results. The baseline UNet showed the lowest performance (Dice = 0.5004, IoU = 0.3356). InceptionV4-UNet achieved the best results (Dice = 0.6400, IoU = 0.4728), followed closely by VGG16-UNet (Dice = 0.6322, IoU = 0.4640).

\begin{figure}[!h]
    \centering
    \includegraphics[width=0.95\linewidth]{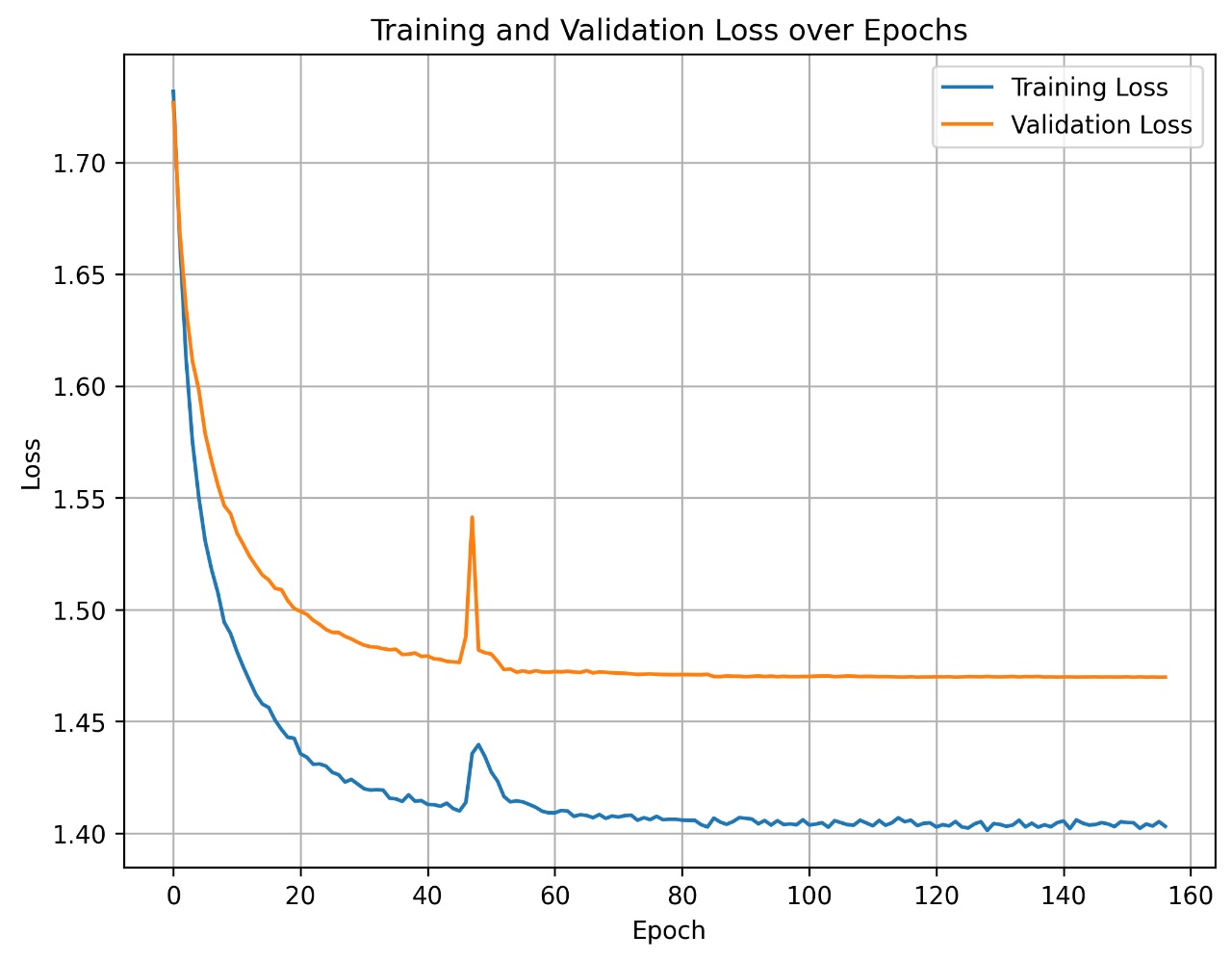}
    \caption{Training and validation loss curves of InceptionV4-UNet during segmentation.}
    \label{fig:inceptionv4unet_loss}
\end{figure}

\begin{table}[!h]
\centering
\caption{Segmentation performance of UNet variants.}
\label{tab:segmentation_results}
\begin{tabular}{cccc}
\hline
\hline
 \textbf{Model} & \textbf{Test Dice} & \textbf{Test IoU} \\
\hline
\hline
 UNet & 0.5004 & 0.3356 \\
 VGG16-UNet & 0.6322 & 0.4640 \\
 InceptionV4-UNet & 0.6400 & 0.4728 \\
 ResNet34-UNet & 0.6040 & 0.4528 \\
\hline
\hline
\end{tabular}
\end{table}
The convergence behavior of the best performing InceptionV4-UNet is illustrated in Figure~\ref{fig:inceptionv4unet_loss}.  
The model demonstrates a steady reduction in training loss with early stabilization of validation loss,  
indicating effective generalization without severe overfitting.   This stable learning curve further supports its superior performance in terms of Dice (0.6400) and IoU (0.4728) compared to other UNet variants.

\subsubsection{Classification:}  
The binary classification results are summarized in Table~\ref{tab:classification_results}. Among the evaluated models, VGG16 achieved the highest accuracy (0.91) and F1 score (0.91), slightly outperforming ResNet50 (Accuracy = 0.86, F1 = 0.86) and InceptionV3 (Accuracy = 0.86, F1 = 0.86). VGG16 also demonstrated superior precision (0.91), indicating stronger reliability in identifying polyp cases without increasing false positives. These results highlight VGG16’s robustness for distinguishing polyp from non-polyp images in this dataset.

\begin{figure*}[!h]
    \centering
    \includegraphics[width=\textwidth]{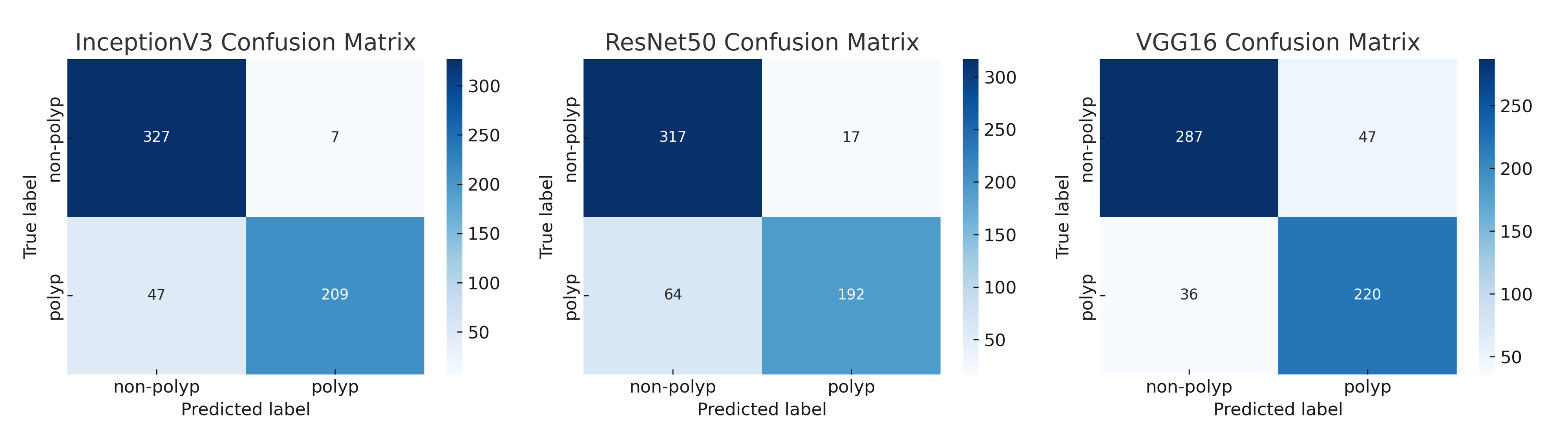}
    \caption{Confusion matrices of InceptionV3, ResNet50 and VGG16  models for binary classification of polyp vs. non-polyp images.}
    \label{fig:confusion_matrices}
\end{figure*}

\begin{table}[!h]
\centering
\caption{Binary classification results on polyp vs. non-polyp images.}
\label{tab:classification_results}
\begin{tabular}{ccccc}
\hline
\hline
\textbf{Model} & \textbf{Accuracy} & \textbf{Precision} & \textbf{Recall} & \textbf{F1 Score} \\
\hline
\hline
VGG16       & 0.9085 & 0.9148 & 0.9085 & 0.9072 \\
ResNet50    & 0.8627 & 0.8689 & 0.8627 & 0.8604 \\
InceptionV3 & 0.8610 & 0.8624 & 0.8610 & 0.8613 \\
\hline
\hline
\end{tabular}
\end{table}
Moreover, the confusion matrices for the three models are shown in Figure~\ref{fig:confusion_matrices}.  
VGG16 demonstrates balanced performance across both classes, correctly identifying 220 polyp images and 287 non-polyp images. ResNet50 shows stronger specificity, with fewer false positives (17 misclassified non-polyp images), but a higher number of missed polyps (64). InceptionV3 achieves the lowest false positive rate (7 non-polyp misclassified), but at the expense of missing 47 polyps. These differences indicate that VGG16 provides the best overall trade-off between sensitivity and specificity, while ResNet50 and InceptionV3 lean towards non-polyp detection strength.

    \section{Discussion}
The benchmarking results highlight the challenges of colonoscopy analysis in resource-constrained environments.  
\par Even with advanced architectures, the best Dice score obtained (0.64 with InceptionV4-UNet) falls well short of values commonly reported for curated datasets such as Kvasir-SEG (>0.85). Figure~\ref{Qual} illustrates this contrast. In successful cases (top row), the predicted masks closely match the ground truth, demonstrating the model’s capacity to segment polyps with clear morphology and favorable lighting. However, in failure cases (bottom row), the model either under-segments polyps or confuses specular highlights and mucosal folds as lesions. These errors occur most frequently in the presence of glare, motion blur, or flat polyps with poor contrast against the background mucosa. Such examples underscore the inherent difficulty of segmentation under real-world imaging artifacts where acquisition protocols, endoscopic equipment, and operator experience vary widely.  
\par  
InceptionV4-UNet achieved the strongest segmentation performance among the tested variants, surpassing both VGG16-UNet and ResNet34-UNet. Its advantage likely stems from the Inception modules’ ability to capture multi-scale contextual information by combining different kernel sizes within each block. This design allows the model to simultaneously encode fine-grained boundaries and larger anatomical structures, making it particularly effective for polyps with irregular morphology. In contrast, VGG16-UNet relies on deep sequential convolutional layers with a higher parameter count, which increases representational power but also makes it more susceptible to overfitting in small or imbalanced datasets. ResNet34-UNet benefitted from residual connections that stabilized training and mitigated vanishing gradients, but its relatively shallower architecture limited its capacity to learn highly diverse features compared to InceptionV4. Overall, while InceptionV4-UNet demonstrated the best generalization in this study, the performance gap to curated benchmarks suggests that segmentation alone cannot yet guarantee reliable polyp delineation in low-resource clinical settings.  
\par  
The classification results complement this picture by showing that simpler transfer learning pipelines can yield more reliable outcomes under the same imaging conditions. VGG16 achieved the highest accuracy (0.91) and F1 score (0.91), outperforming ResNet50 and InceptionV3. Importantly, its precision (0.91) indicates strong reliability in identifying polyp frames without increasing false positives, which is critical for reducing unnecessary clinical alarms. ResNet50 and InceptionV3 performed competitively but leaned toward higher specificity, missing more true polyp instances in the process. These findings reveal a key distinction: while segmentation models struggled to delineate lesion boundaries consistently, classification models maintained stable performance in simply identifying the presence or absence of polyps.  
\par  
Taken together, these results suggest that a hybrid approach may provide the most practical solution in resource-limited environments. A lightweight classification model, such as VGG16, could first screen video streams to flag candidate frames containing polyps. Subsequently, a segmentation model (e.g., InceptionV4-UNet) could be applied selectively to localize and highlight suspicious regions. This tiered approach would reduce computational overhead, limit false alarms, and improve interpretability for clinicians. Furthermore, it mirrors real-world diagnostic workflows where a physician first identifies a suspicious frame before carefully assessing lesion boundaries.  
\par  
Despite these promising directions, several limitations remain. First, performance across all models was constrained by the variability and noise inherent in low-resource colonoscopy data. Common failure modes—such as specular highlights misclassified as lesions, motion artifacts, and poor mucosal contrast—highlight the need for preprocessing pipelines tailored to real-world acquisition conditions. Second, while InceptionV4-UNet showed improved generalization, its training complexity and resource requirements may limit deployment in hospitals without high-end hardware. Finally, the lack of external validation on independent datasets restricts the generalizability of the findings. Future work should therefore focus on multi-center datasets, domain adaptation strategies, and lightweight architectures optimized for edge devices to enable widespread clinical use.  
\par  
In summary, segmentation in real-world colonoscopy remains an open challenge, but classification results demonstrate clear potential for reliable polyp detection in low-resource environments. By integrating classification for frame selection and segmentation for localization, a clinically viable CAD pipeline can be constructed, offering a balanced compromise between accuracy, computational feasibility, and clinical usability in under-resourced hospitals.

\section{Limitations}

This study has several limitations that should be acknowledged. First, although all polyp masks were reviewed by medical professionals, the initial annotations were performed by trained non-clinical annotators, which may introduce subtle inconsistencies compared to expert-only annotations. Second, the dataset lacks detailed patient demographic information (e.g., age, sex, clinical history), preventing analysis of how polyp appearance may vary across patient groups. Third, while the dataset size is valuable in a low-resource context, it remains modest compared to large international repositories, which may limit the diversity of polyp morphologies captured. Finally, the study represents data from a single public hospital in Bangladesh, and therefore the findings may not fully generalize to other healthcare settings with different equipment, patient populations, or clinical protocols.

    \section{Conclusions}
In this study, we introduced the BUET Polyp Dataset, a collection of polyp and non-polyp colonoscopy images acquired under real-world, resource-constrained conditions. By benchmarking both segmentation and classification models, we systematically evaluated the challenges posed by imaging artifacts such as glare, motion blur, and uneven illumination. Our results demonstrate that segmentation remains particularly difficult in these settings: even advanced UNet variants with pretrained backbones achieved modest Dice scores ($\approx 0.64$), substantially lower than those reported on curated datasets. These findings emphasize the limitations of deploying models trained solely on idealized data and highlight the pressing need for datasets that capture realistic clinical variability. In contrast, binary classification using transfer learning showed stronger performance, with VGG16 achieving the highest accuracy (0.91). This suggests that classification-based pipelines may serve as reliable first-stage detectors in clinical computer-aided diagnosis (CAD) systems, while segmentation models require further refinement to handle the complexities of low-resource imaging. Future directions include expanding the dataset across multiple hospitals, incorporating temporal information from video data, and exploring domain adaptation or self-supervised pretraining approaches. These efforts will be crucial to develop robust, generalizable AI systems capable of assisting gastroenterologists in improving colorectal cancer screening outcomes worldwide.

\section*{CRediT authorship contribution statement}
\noindent\textbf{Ridoy Chandra Shil:} Data curation, Software, Validation, Writing – review \& editing.
\newline
\textbf{Ragib Abid:} Data curation, Formal analysis, Visualization, Writing – original draft. , Writing – review \& editing.  
\newline
\textbf{Tasnia Binte Mamun:}  Formal analysis, Visualization, Writing – original draft. , Writing – review \& editing.  
\newline
\textbf{Samiul Based Shuvo:} Conceptualization, Methodology, Supervision, Validation, Writing – original draft, Writing – review \& editing. 
\newline
\textbf{Masfique Ahmed Bhuiyan:} Data curation, Project administration, Writing – review. 
\newline
\textbf{Jahid Ferdous:} Supervision, Project administration, Writing – review \& editing, Funding acquisition.

\section*{Declarations}

\subsection*{Ethical approval}
This study was conducted on retrospectively collected, anonymized, and randomized patient data. No identifiable patient information was used, and, in accordance with institutional guidelines, formal ethical approval was not required for this retrospective analysis.

\subsection*{Consent to participate}
Not applicable, as this study involved retrospective anonymized data.

\subsection*{Consent to publish}
Not applicable.

\subsection*{Funding}
None.

\subsection*{Conflict of interest}
The authors declare that they have no conflict of interest.

\subsection*{Use of Generative AI}
Artificial intelligence tools were used only to improve clarity, grammar, and linguistic expression. These tools were not involved in generating scientific content, data interpretation, or analysis. The authors remain fully responsible for the originality, accuracy, and integrity of all intellectual content.

\bibliography{mybibfile}

\end{document}